\documentclass[12pt,tightenlines,showkeys]{revtex4}
\pdfoutput=1

\usepackage{graphicx}
\usepackage{amsmath}
\usepackage{amssymb}
\usepackage{bm}
\usepackage{color}
\usepackage[loose]{subfigure}

\graphicspath{{figs/}{figs_lo/}}

\begin{document}

\title{Mixing effectiveness depends on the source-sink structure:
  Simulation results}

\author{Takahide Okabe}
\affiliation{Department of Physics and Institute for Fusion Studies,
  The University of Texas at Austin, Austin, TX 78712-0264, USA}

\author{Bruno Eckhardt}
\affiliation{Fachbereich Physik Philipps-Universtit\"at, D-35032
  Marburg, Germany}

\author{Jean-Luc Thiffeault}        
\affiliation{Department of Mathematics, University of Wisconsin,
  Madison, WI 53706-1388, USA}

\author{Charles R. Doering}
\affiliation{Departments of Mathematics and Physics, University of
  Michigan, Ann Arbor, MI 48109-1043, USA}

\date{\today}

\begin{abstract}
  The mixing effectiveness, i.e., the enhancement of molecular diffusion,
  of a flow can be quantified in terms of the suppression of concentration
  variance of a passive scalar sustained by steady sources and sinks.
  The mixing enhancement defined this way is the ratio of the RMS  
  fluctuations of the scalar mixed by molecular diffusion alone to
  the (statistically steady-state) RMS fluctuations of the scalar density
  in the presence of stirring.  This measure of the
  effectiveness of the stirring is naturally related to the
  enhancement factor of the equivalent eddy diffusivity over molecular
  diffusion, and depends on the P\'eclet number. 
  It was recently noted that the maximum possible
  mixing enhancement at a given P\'eclet number depends as well on the
  structure of the sources and sinks.  That is, the mixing efficiency, the
  effective diffusivity, or the eddy diffusion of a flow generally depends
  on the sources and sinks of whatever is being stirred.  
  Here we present the results of particle-based simulations quantitatively
  confirming the source-sink dependence of
  the mixing enhancement as a function of P\'eclet number for a model
  flow.
\end{abstract}

\keywords{
    stirring \& mixing;
    transport processes;
    stochastic particle dynamics;
    Brownian motion;
    turbulence}

\maketitle

\section{Introduction}

Mixing by fluid flows is a ubiquitous natural phenomenon that plays a
central role in many of the applied sciences and engineering.  A
geophysical example is the mixing of aerosols (e.g., $\text{CO}_2$
supplied by a volcano, say, or by human activity) in the atmosphere.
Aerosols are dispersed by molecular diffusion on the smallest scales
but are more effectively spread globally by atmospheric flows.  The
density --- and density fluctuations --- of some aerosols influence
the albedo of the earth and thus have a significant environmental
impact.  Hence it is important to understand fundamental properties of
dispersion, mixing, and the reduction of concentration fluctuations
by stirring flow fields.

Various aspects of mixing have been the
focus of many review articles~\cite{Ottino1990,Majda1999,Warhaft2000,%
Shraiman2000,Sawford2001,Falkovich2001,Aref2002,Wiggins2004}.
At the most basic level, the mixing of a passive scalar can be modeled
by an advection-diffusion equation for the scalar concentration field
with a specified stirring flow field.  In this work we will focus on
problems where fluctuations in the scalar field are generated and
sustained by temporally steady but spatially inhomogeneous sources.
The question of interest here is this: for a given source
distribution, how well can a specified stirring flow mix the scalar
field?  

Mixing effectiveness can be measured by the scalar variance over the domain.
A well-mixed scalar field will have a more uniform density with
relatively ``small'' variance while increased fluctuations in the scalar density
will be reflected in a ``large'' variance.  We put quotes around the
quantifiers small and large because the variance is a dimensional
quantity that needs an appropriate dimensional point of reference 
from which it is being measured.

Several years ago Thiffeault {\it et al} \cite{Thiffeault2004} introduced a
notion of ``mixing enhancement'' for a velocity field stirring a steadily
sustained scalar by comparing the bulk (space-time) averaged density
variance with and without advecting flow.  Mixing is accomplished by 
molecular diffusion alone in the absence of stirring, which can be
quite effective on small scales but is not generally so good at
breaking up and dispersing large-scale fluctuations quickly.  
Stirring can greatly enhance the transport of the scalar from regions
of excess density to depleted regions, suppressing the variance far
below its diffusion-only value.  The magnitude of this variance
suppression by the stirring --- the ratio of the variance without
stirring to the variance in the presence of stirring --- is a
dimensionless quantity that provides a sensible gauge of the mixing
effectiveness of the flow.  Different advection fields will have different
mixing efficiencies stirring scalars supplied by different sources.  It is
then of obvious interest both to determine theoretical limits on
mixing enhancements for various source configurations and to explore
whether those limits may be approached---or even perhaps achieved----for 
particular flows.

There have been many studies of stirring and mixing of a scalar with
fluctuations sustained by spatially inhomogeneous sources and sinks.  Some of the
earliest are by Townsend~\cite{Townsend1951,Townsend1954}, who was
concerned with the effect of turbulence and molecular diffusion on a
heated filament.  He found that the spatial localization of the source
enhanced the role of molecular diffusivity.  Durbin~\cite{Durbin1980}
and Drummond~\cite{Drummond1982} introduced stochastic particle models
to turbulence modeling, and these allowed more detailed studies of the
effect of the source on diffusion.  Sawford and
Hunt~\cite{Sawford1986} pointed out that small sources lead to a
dependence of the variance on molecular diffusivity.  These models
were further refined by~\cite{Thomson1990,Borgas1994,Sawford2001}.
Chertkov {\it et
  al}~\cite{Chertkov1995,Chertkov1995b,Chertkov1997,Chertkov1997b,%
  Chertkov1998} and Balkovsky \& Fouxon~\cite{Balkovsky1999}
  addressed the case of a random, statistically-steady source.

In this paper we study the enhancement of mixing by an advection field
using a particle-based computational scheme that is easy to implement
and applicable to a variety of source distributions.
The idea is to develop a method
that accurately simulates advection and diffusion of large numbers of
particles supplied by a steady source, and to measure density
fluctuations by ``binning'' the particles to produce
an approximation of the hydrodynamic concentration field.  Unlike a
numerical PDE code, a particle code does not prefer specific forms of
the flow or the source (PDE methods generally work best with 
very smooth fields).  
There is, however, no free lunch: the accuracy of the
particle code is ultimately limited by the finite number of particles
that can be tracked.  The limitation to finite numbers of particles
inevitably introduces statistical errors due to discrete fluctuations
in the local density and systematic errors in the variance
measurements due to binning.  But these problems are tractable, and as
we will show, the method proves to be quantitatively accurate and 
computationally efficient for some applications.

\section{Theoretical background}

In this section we review basic facts about the mixing enhancement
problem as formulated by Thiffeault, Doering \& Gibbon {\it et al}
\cite{Thiffeault2004} and developed by Plasting \& Young \cite{Plasting2006},
Doering \& Thiffeault \cite{DoeringThiffeault2006}, Shaw {\it et al} \cite{Shaw2007}, and
Thiffeault and Pavliotis \cite{Thiffeault2008}.  The dynamics is given
by the advection-diffusion equation for the concentration of a passive
scalar $\rho(t, \bm{x})$ with time-independent but spatially
inhomogeneous source field $S(\bm{x})$:
\begin{equation}
\frac{\partial\rho}{\partial t}+\bm{u}\cdot\nabla\rho=\kappa \Delta \rho + S(\bm{x}),
\label{ADE1}
\end{equation}
where $\kappa$ is the molecular diffusivity and $\bm{u}(t,
\bm{x})$ is a specified advection field that satisfies (at each
instant of time) the incompressibility condition
\begin{equation}
\nabla \cdot \bm{u} = 0.
\end{equation}
For simplicity, the domain is the $d$-torus, i.e., $[0, L]^d$ with
periodic boundary conditions.  
We limit attention to stirring
fields that satisfy the properties of statistical homogeneity and
isotropy in space defined by
\begin{equation}
\overline{u_{i}(\cdot, \bm{x})} = 0,\qquad
\overline{u_{i}(\cdot, \bm{x})u_{j}(\cdot, \bm{x})} = \frac{U^{2}}{d}\delta_{ij}
\end{equation}
where the overbar denotes time-averaging and $U$ is the root mean
square speed of the velocity field, a natural indicator of the
intensity of the stirring.  
These are statistical properties of
homogeneous isotropic turbulence on the torus,
but they are also shared by many other kinds of flows.

We are interested in fluctuations in the concentration $\rho$ so the
spatially averaged background density is irrelevant.  It is easy to
see from \eqref{ADE1} that the spatial average of $\rho$ grows
linearly with time at the rate given by the spatial average of $S$.
Hence we change variables to spatially mean-zero quantities
\begin{equation}
 \theta (t, \bm{x}) =  \rho(t, \bm{x}) - \frac{1}{L^d}\int \mathrm{d}^{d}x'\, \rho(t, \bm{x}')
\end{equation} 
and
\begin{equation}
s(\bm{x}) = S(\bm{x}) - \frac{1}{L^d}\int \mathrm{d}^{d}x'\, S(\bm{x}')
\end{equation} 
that satisfy
\begin{equation}
\frac{\partial\theta}{\partial t}+\bm{u}\cdot\nabla\theta=\kappa \Delta \theta + s(\bm{x}).
\label{ADE2}
\end{equation}
(We must also supply initial conditions for $\rho$ and/or $\theta$ but
they play no role in the long-time steady statistics that we are
interested in.)

The ``mixedness'' of the scalar may be characterized by, among other
quantities, the long-time averaged variance of $\rho$, proportional to
the long-time averaged $L^2$ norm of $\theta$,
\begin{equation}
\langle \theta^2 \rangle:=\lim_{T \to \infty}\frac{1}{T} \int_{0}^{T}\mathrm{d}t \
\frac{1}{L^d}\int \mathrm{d}^dx\, \theta^2(t, \bm{x})
\end{equation}
The smaller $\langle \theta^2 \rangle$ is, the more uniform the
distribution.  The ``mixing enhancement'' of a stirring field is
naturally measured by comparing the scalar variance to the variance
with the same source but in the absence of stirring.  To be precise,
we compare $\langle \theta^2 \rangle$ to $\langle \theta_{0}^{\ 2}
\rangle$ where $\theta_{0}$ is the solution to
\begin{equation}
\frac{\partial\theta_{0}}{\partial t}=\kappa \Delta \theta_{0} + s(\bm{x})
\label{ds}
\end{equation}
(with, say, the same initial data although these will not affect the
long-time averaged fluctuations).  
The dimensionless {\it mixing enhancement factor} is then defined as
\begin{equation}
{\cal E}_{0}:=\sqrt{\frac{\langle \theta_{0}^{\,2} \rangle}{\langle \theta^2 \rangle}}.
\end{equation} 
This quantity carries the subscript $0$ because we can also define
{\it multiscale mixing enhancements} \cite{DoeringThiffeault2006,Shaw2007} by weighting
large/small wavenumber components of the scalar fluctuations:
\begin{equation}
{\cal E}_{p}:=\sqrt{\frac{\langle |\nabla^{p}\theta_{0}|^{2} \rangle}{ \langle|\nabla^{p}\theta|^{2} \rangle}}\,,
\qquad p=-1,0,1.
\end{equation}
As discussed in Doering \& Thiffeault \cite{DoeringThiffeault2006}, Shaw {\it et al}
\cite{Shaw2007} and Shaw \cite{ShawGFD2005}, ${\cal E}_{\pm1}$ provide a gauge of the
mixing enhancement of the flow as measured by scalar fluctuations on
relatively small and large length scales, respectively. 
We refer to them as enhancement factors because if one were to define an
effective, eddy, or equivalent diffusivity $\kappa_{e,p}$ as the value of a molecular
diffusion necessary to produce the same value of $\langle |\nabla^{p}\theta|^{2} \rangle$
with stirring, then $\kappa_{e,p}=\kappa {\cal E}_{p}$.
 In this paper, however, we will focus exclusively on ${\cal E}_{0}$, the mixing enhancement at
``moderate'' length scales.

There is a theoretical upper bound on ${\cal E}_{0}$ valid for any
statistically stationary homogeneous and isotropic stirring field
\cite{DoeringThiffeault2006, ShawGFD2005, Shaw2007}:
\begin{equation}
{\cal E}_{0} \leq \sqrt{\frac{\sum_{\bm{k} \neq \bm{0}}|\hat{s}(\bm{k})|^{2}/k^{4}}{\sum_{\bm{k} \neq \bm{0}}|\hat{s}(\bm{k})|^{2}/(k^{4}+\frac{{\rm Pe}^{2}}{L^{2}d}k^{2})}}
\label{bound}
\end{equation} 
where $\hat{s}(\bm{k})$ are the Fourier coefficients of the source
and the P\'eclet number,
\begin{equation}
{\rm Pe}:={UL}/{\kappa},
\end{equation}
is a dimensionless measure of the intensity of the stirring.
Generally, we anticipate that ${\cal E}_{0}$ is an increasing function
of Pe and the estimate in \eqref{bound} guarantees that 
${\cal E}_{0}(\text{Pe}) \lesssim \text{Pe}$ as $\text{Pe} \rightarrow
\infty$, the ``classical'' scaling necessary if there is to be any residual variance
suppression in the singular vanishing diffusion limit.
That is, if ${\cal E}_{0}(\text{Pe}) \sim \text{Pe}$ then
$\kappa_{e,0}$ has a nonzero limit as $\kappa \rightarrow 0$
with all other parameters held fixed.
It is natural to refer to any of the possible sub-classical scalings as ``anomalous''.

The upper limit to the mixing enhancement in \eqref{bound} depends on the
stirring field only through $U$ via Pe, but it depends on all the
details of the source distribution.  As studied in depth in references
\cite{DoeringThiffeault2006,Shaw2007,ShawGFD2005}, the structure of the scalar source can have a
profound effect on the high Pe scaling of ${\cal E}_{0}$, notably for sources with small scales.
It is physically meaningful to consider measure-valued source-sink distributions, like delta-functions,
with arbitrarily small scales.  
It is precisely this source size dependence of ${\cal E}_{0}(\text{Pe})$ that
motivates the development of a computational method that can handle
singular source distributions.

In this study, for computational simplicity and efficiency, we utilize the ``random sine flow'' as the stirring field.
In the two-dimensional case this is defined for all time by
\begin{equation}
{\bm{u}}(t,\bm{x})= \begin{cases}
w\sin \left({2 \pi y}/{L}+\phi\right)\hat{\textbf{\i}}\,, \qquad &nT< t \le nT+\tfrac12T;\\
w\sin \left({2 \pi x}/{L}+\phi'\right)\hat{\textbf{\j}}\,, \qquad
& nT+\tfrac12T < t \le (n+1)T,
\end{cases}
\end{equation}
where $T$ is the period, $n=0, 1, 2,\dots$, and $\phi$ and $\phi'$
are random phases chosen independently and uniformly on $[0,2\pi)$ in
each half cycle, which assures the homogeneity of the flow field. In
this case, $w=\sqrt{2}U$. 
In the three-dimensional case, we employ
\begin{equation}
\bm{u}(t, \bm{x})=
\begin{cases}
w\left[\sin \alpha \sin\left(2\pi y/L+\phi_{2}\right)+\cos \alpha \sin\left(2 \pi z/L+\phi_{3}\right) \right]\hat{\textbf{\i}}\,,\  &nT < t \leq nT+ \tfrac13T; \nonumber \\
w\left[\sin \alpha \sin\left(2 \pi z/L+\psi_{3}\right)+\cos \alpha \sin\left(2 \pi x/L+\phi_{1}\right) \right]\hat{\textbf{\j}}\,,\ &nT+\tfrac13T < t \leq nT+\tfrac23T; \nonumber \\
w\left[\sin \alpha \sin\left(2 \pi x/L+\psi_{1}\right)+\cos \alpha \sin\left(2 \pi y/L+\psi_{2}\right) \right]\hat{\textbf{k}}\,,\ &nT+\tfrac23T < t \leq (n+1)T, \nonumber 
\end{cases}
\end{equation}
where again $w=\sqrt{2}U$, $n=0,1,2,\dots$, and $\alpha$,
$\phi_{1,2,3}$ and $\psi_{1,2,3}$ are uniform random numbers in $[0,
2\pi)$ chosen independently every $T/3$.  The angle~$\alpha$ randomizes
the shear direction to guarantee isotropy of the flow.

\section{Numerical method}

In a particle code for solving the advection-diffusion equation, the
concentration field $\rho$ is represented by a distribution of
particles.  Particles are introduced by generating random locations
using the properly normalized source $S(\bm{x})$ as a probability
distribution function, then they are transported by advection and
diffusion.  The particle density, $\rho(t, \bm{x})$, is
measured by covering the domain with bins counting the number of
particles per bin.

A discrete particle method is employed because it can easily deal with small-scale
sources such as $\delta$ functions.  
It is also straightforward to implement with any advection field.  
The downside of a particle method is that it necessarily involves two kinds of errors: 
the number density of particles calculated by dividing the domain into bins is
only resolved down to the size of the bins, and the
measurement of $\rho$ always includes statistical errors due to
the use of finite numbers of particles.

\subsection{Time evolution}

At each time step the system is evolved by advection, diffusion, the source, and sinks.  
An advection-only equation would be solved by moving particles along characteristics, 
and a diffusion-only equation would be solved by adding independent Gaussian noises
to each coordinate of each particle.  
With both advection and diffusion we need to solve a stochastic differential equation to 
determine the proper displacement of the particles during a time step.  
The stochastic differential equation is
\begin{equation}
d \bm{X}=\bm{u}(t, \bm{X})dt+\sqrt{2\kappa}\ d\bm{W}
\label{eq:sde}
\end{equation}
where $\bm{W}(t)$ is a standard vector-valued Wiener process. 

In order to solve \eqref{eq:sde}, we will consider cases where the displacement
due to the noise in a subinterval of length $T/d$ (where $d$ is the dimension) 
is much smaller than the wavelength of the random sine flow.  
This condition is realized better and better as Pe increases.
Then, during each subinterval, the drift field $\bm{u}(t,
\bm{X})$ experienced by each particle can be approximated by a
steady flow with a linear shear.  
In 2D, for the first
half of the period for a particle starting at $(x_{0}, y_{0})=(X(t=0),
Y(t=0))$ we approximate \eqref{eq:sde} by
\begin{align}
\begin{split}
dX&=w\sin\left({2\pi y_{0}}/{L}+\phi\right)dt+w\cos\left({2 \pi y_{0}}/{L}+\phi\right)\dfrac{2 \pi}{L}(Y-y_{0})dt+\sqrt{2\kappa}\,dW_{1},\\
dY&=\sqrt{2\kappa}\,dW_{2},
\end{split}
\end{align}
and for the second half of the period, starting from $(x'_0,y'_0)=(X(t=T/2), Y(t=T/2))$,
\begin{align}
\begin{split}
dX&=\sqrt{2\kappa}\,dW_{1}, \\
dY&=w\sin\left({2\pi x'_{0}}/{L}+\phi '\right)dt+w\cos\left({2 \pi x'_{0}}/{L}+\phi '\right)\dfrac{2 \pi}{L}(X-x'_{0})dt+\sqrt{2\kappa}\,dW_{2}\,.
\end{split}
\end{align}
Therefore, during the first half period we evolve the position of a
particle through a time interval $\Delta t$ (where $\Delta t \le T/2$
need {\it not} be small) by the map
\begin{align}
\begin{split}
x_{0} &\rightarrow x_{0}+w\sin\left({2\pi y_{0}}/{L}+\phi\right)\Delta t+ R_{1}, \\
y_{0} &\rightarrow y_{0}+R_{2},
\end{split}
\end{align}
where $R_{1}$ and $R_{2}$ satisfy 
\begin{align}
\begin{split}
dR_{1}&= S_{2}R_{2}dt+\sqrt{2 \kappa}dW_{1} \qquad \left(S_{2}:=2\pi w
  L^{-1}\cos\left({2 \pi y_{0}}/{L}+\phi\right)\right), \\
dR_{2}&=\sqrt{2\kappa}dW_{2}.
\end{split}
\end{align}
The variance-covariance matrix of $R_{1}$ and $R_{2}$ is
\begin{equation}
 \left(
\begin{array}{cc}
\bm{E}(R_{1}^{\ 2})& \bm{E}(R_{1}R_{2})  \\
\bm{E}(R_{2}R_{1})& \bm{E}(R_{2}^{\ 2})  \\
\end{array}
\right)
=
 \left(
\begin{array}{cc}
\frac{2}{3}S_{2}^{\ 2}\kappa t^3+2\kappa t& S_{2}\kappa t^2\\
S_{2} \kappa t^2&  2\kappa t \\
\end{array}
\right),
\label{eq:varcovar}
\end{equation}
which is realized by 
\begin{align}
R_{1}&=\sqrt{\tfrac{1}{6}S_{2}^{\ 2}\kappa t^3+2\kappa t}\times N_{1}+\sqrt{\tfrac{1}{2}S_{2}^{\ 2}\kappa t^3} \times N_{2},\\
R_{2}&=\sqrt{2\kappa t}\times N_{2},
\end{align}
where $N_1$ and $N_2$ are independent $N(0,1)$ random variables
(normally distributed with mean~$0$ and standard deviation~$1$).  The
matrix~\eqref{eq:varcovar} describes the evolution of a passive scalar
field in a shear flow~\cite{Taylor1954,Aris1956,Rhines1983}.

\begin{figure}
\begin{center}
\subfigure[]{
  \includegraphics[height=4.15cm, angle=270]{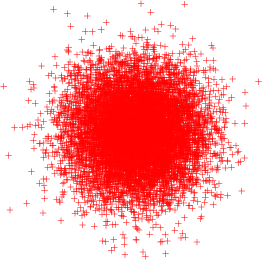}
  \label{fig:gauss1}
}\hspace{3em}%
\subfigure[]{
  \includegraphics[height=5cm, angle=270]{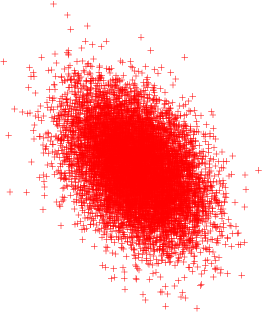}
  \label{fig:gauss2}
}
\end{center}
\caption{(a) A circular Gaussian distribution of particles transported
  and sheared into (b) an elliptical Gaussian cloud.}
\label{fig:gauss}
\end{figure}

Therefore the time evolution map during the first half period is
\begin{subequations}
\begin{align}
x_0 &\rightarrow x_0+w\sin\left({2\pi y_{0}}/{L}+\phi\right)\,\Delta t+
\sqrt{\tfrac{1}{6}S_2^{\ 2} \kappa (\Delta t)^3 + 2 \kappa \Delta t} \, N_1 +
 \sqrt{\tfrac{1}{2} S_2^{\ 2} \kappa (\Delta t)^3} \, N_2\,, \label{eq:evolution1}\\
y_0 &\rightarrow y_0+\sqrt{2 \kappa \Delta t} \, N_2\,.  \label{eq:evolution2} 
\end{align}%
\end{subequations}%
A similar map is employed during the second half of the period.  These
stochastic maps include the shear --- in the approximation that the
shear remains constant for each particle during each half cycle ---
that causes a ``distortion'' of a Gaussian cloud of particles; see Fig.~\ref{fig:gauss}.

The same calculations apply for the three-dimensional case: for the
first subinterval, the time-evolution map is
\begin{subequations}
\begin{align}
x_0 &\rightarrow x_0+w\left[\sin \alpha \sin\left({2\pi y_0}/{L}+\phi_{2}\right)+\cos \alpha \sin\left({2 \pi z_0}/{L}+\phi_{3}\right) \right]\nonumber\\
&\phantom{\rightarrow x_0}+\sqrt{\tfrac{1}{6}(S_{2}^{\ 2}+S_{3}^{\ 2})\kappa (\Delta t)^{3}+2\kappa\Delta t } \, N_{1}
+S_{2}\sqrt{\tfrac12\kappa}(\Delta t) ^{\frac{3}{2}} \, N_{2}
+S_{3}\sqrt{\tfrac12\kappa}(\Delta t) ^{\frac{3}{2}} \, N_{3},\\
y_0 &\rightarrow y_0+\sqrt{2 \kappa \Delta t} \, N_{2},\\
z_0 &\rightarrow z_0+\sqrt{2 \kappa \Delta t} \, N_{3},
\end{align}
\end{subequations}
where $S_{2}= 2\pi w L^{-1} \sin\alpha
\cos({2\pi y_{0}}/{L}+\phi_{2})$, $S_{3}= 2\pi w L^{-1} \cos\alpha
\cos({2\pi z_{0}}/{L}+\phi_{3})$, and $N_1, N_2$ and $N_{3}$ are
independent $N(0,1)$ random variables. The maps for the other
subintervals can be obtained by cyclic permutation of the coordinates.

The steady scalar source is realized by introducing a new particle one
by one using normalized $S(\bm{x})$ as a probability distribution
function.  Numerically, such a probability distribution function can
be realized by mapping uniform random numbers over $[0,1]$ with the
inverse of the cumulative probability distribution function in
question.

New particles are added constantly so the total number of particles
continues to increase, which slows down the computation.  To cope with
increasing particles, we implement a particle subtraction scheme.
Particles eventually get well mixed and ``older'' particles do not
contribute to the value of the hydrodynamic variance.  There is no
added value in keeping track of particles that have been in the mix
for a very long time, and we can simply remove them from the system
after a sufficiently long time.  It is very important to keep track of
the ``age'' of each particle, however, and to only remove sufficiently
old well-mixed particles.  (For example, if a random fraction of
particles is removed at regular time intervals, then the simulation
becomes one of a system particles with a random finite lifetime,
described by an advection-diffusion equation with an additional
density decay term.)

In order to determine how old particles must be in order to safely
remove them without affecting the hydrodynamic variance, prior to a
full simulation run a test is performed as follows.  Starting from an
initial set of $N_{i}$ particles located in space according to the
source distribution, the flow and diffusion are allowed to act and the
variance of the number of particles per bin, which decays with time,
is monitored.  The number $N_{i}$ is of the order of the number of
particles that are introduced in the full simulation during, say, an
interval of length $T$ characteristic of the random sine flow.  The
variance does not decay all the way to zero, however, but rather to
the variance expected when $N_{i}$ particles are randomly distributed
among the bins.  The time when the variance achieves this
random-distribution variance, measured beforehand for a given flow and
diffusion strength, is then the required ``aging'' time before
particles can be safely removed in the full simulation with the steady
source.  Such a trial run is performed for each flow, diffusion
strength, source distribution and particle number because this
``mixing time'' depends on all these factors.  Futher details of the
criteria for removing old particles and extensive tests and benchmark
trials may be found in Ref.~\cite{OkabeGFD2006}.

\subsection{Variance calculation and background noise}

The variance $\langle \theta^2\rangle$ is measured by monitoring the
fluctuations in the number of particles per bin, and time-averaging.
In $d$ dimensions the domain is divided into $l^d$ bins and the code
calculates $\langle n^2 \rangle$, where $n$ is the number of particles
in a bin.
Then $\langle \theta^2\rangle$ is initially approximated by
\begin{equation}
  \langle n^2 \rangle - \langle n \rangle^2
  = \left(L/l\right)^{2d} \langle \theta^2 \rangle.
  \label{measure}
\end{equation}
We say ``initially'' because the expression above includes both the
hydrodynamic fluctuations of interest {\it and} discreteness
fluctuations resulting solely from the fact that each bin contains a
finite number of particles.

The subtraction scheme eliminates the ``well-mixed'' particles that do
not contribute to the value of the hydrodynamic variance.  But even if
the system were completely mixed so that theoretically, $\langle
\theta^2 \rangle = 0$, the measured variance $\langle n^2 \rangle -
\langle n \rangle^2$ would be (very close to, for small bins) $\langle
n \rangle$, which is on the order of $N/l^d$, where $N$ is the total
number of particles in the domain.  This follows from the fact that
$\theta(t, \bm{x})$ is represented in this particle method by only a
finite number of particles in each finite size bin.  That is,
$\langle \theta^2 \rangle$ as defined by (\ref{measure}) is nonzero
even when the particles are uniformly distributed: then the bulk
variance includes fluctuations as if $N$ particles were randomly
thrown in $l^d$ bins.  The helpful fact is that the bulk variance
contribution from these {\it background fluctuations} due to finite
numbers of particles in the bins does not depend on (i.e., is
uncorrelated with) the hydrodynamic density variation from bin to bin.
The total contribution to the variance is the sum of the ``extra''
variance in each bin which is linear in the (mean) number of particles
in each bin.  Hence the sum of the variances is $\sim N$ and the bulk
variance contribution from the background fluctuations,
$Nl^{d}/L^{2d}$, can simply be subtracted from the initial estimate
for $\langle \theta^2 \rangle$ in (\ref{measure}).  The net result is
our measured value of the hydrodynamic variance.

In addition to the inevitable fluctuations due to discreteness,
density variations are observed only down to the length scales $\sim L/l$
because of the binning density, which is another source of error in
this procedure.
We use $l \ge 100$, which tests and benchmark studies indicate
is sufficient for the examples studies here~\cite{OkabeGFD2006}.

The variance is calculated once for each subinterval, and the instant
when it is calculated is determined randomly in order to obtain an
unbiased time average.  Thus each subinterval is divided into two
parts, before and after variance calculation, and the particle
transport and source processes are appropriately adapted.
The final measured quantities are long time averages that are 
observed to be converged to within the error indicated on the plots below.

\section{Results}

\begin{figure}
\begin{center}
\subfigure[\ {$a=L/2$}]{
  \includegraphics[height=5cm, angle=270]{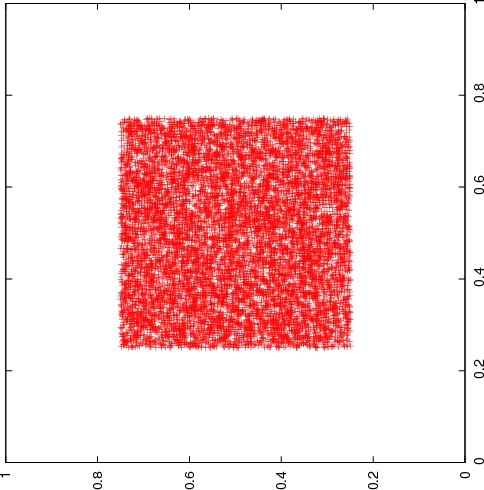}
  \label{fig:sqfig2}
}\hspace{3em}%
\subfigure[\ {$a=L/10$}]{
  \includegraphics[height=5cm, angle=270]{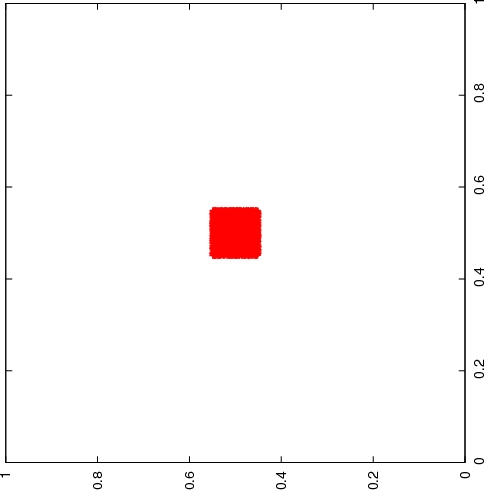}
  \label{fig:sqfig10}
}
\caption{Square-shaped source of two different sizes; particles shown
sampled from the uniform source distribution over the squares.}
\label{fig:sqfig}
\end{center}
\end{figure}

In order to investigate the effect of source-sink scales on maximal
and actual mixing enhancements, we performed a series of
simulations for square-shaped sources of various sizes $a < L$ as
illustrated in Fig.~\ref{fig:sqfig}.

\begin{figure}
\begin{center}
\subfigure[\ {$d=2$}]{
  \includegraphics[height=6cm]{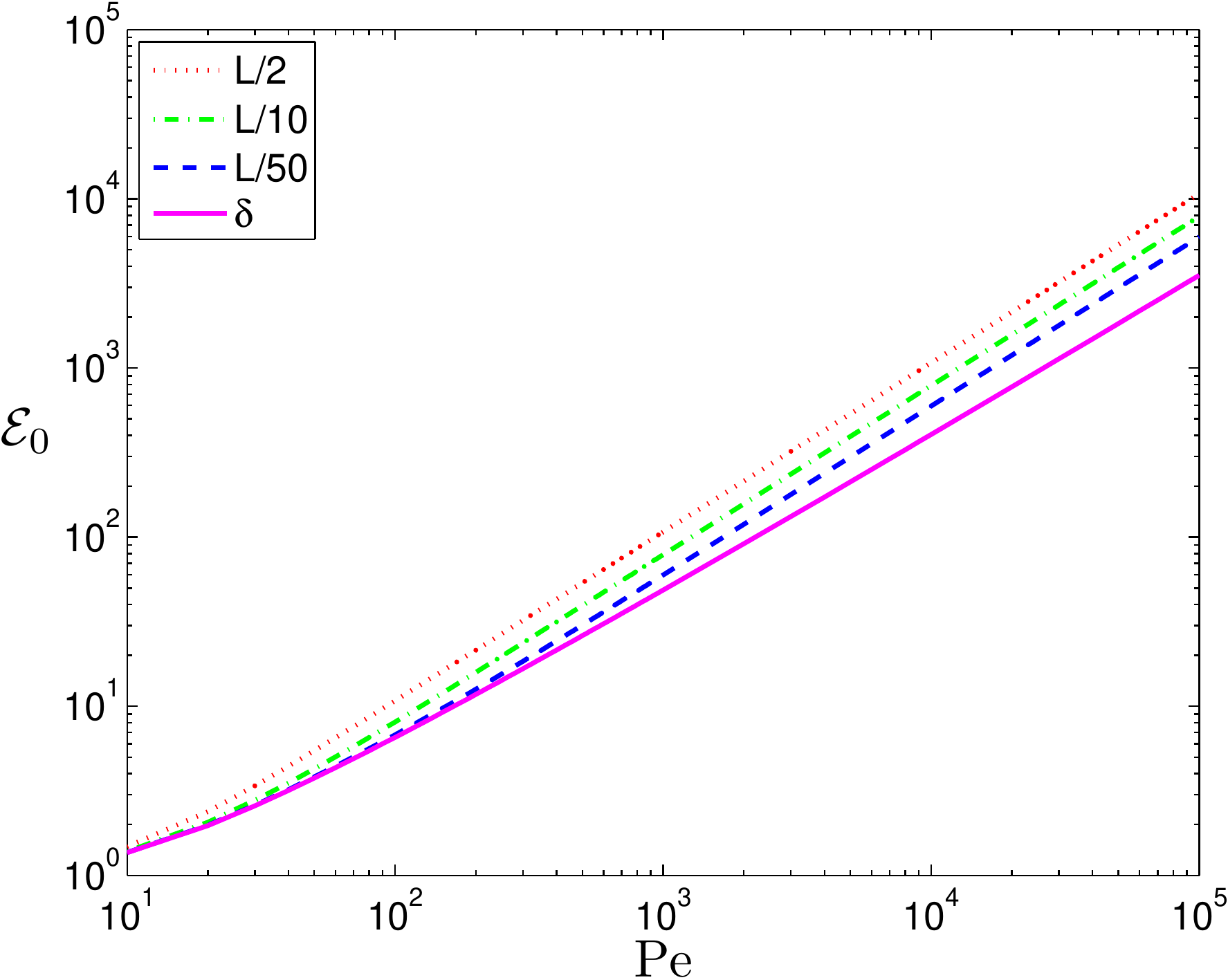}
  \label{fig:2Dbounds}
}\hspace{2em}%
\subfigure[\ {$d=3$}]{
  \includegraphics[height=6cm]{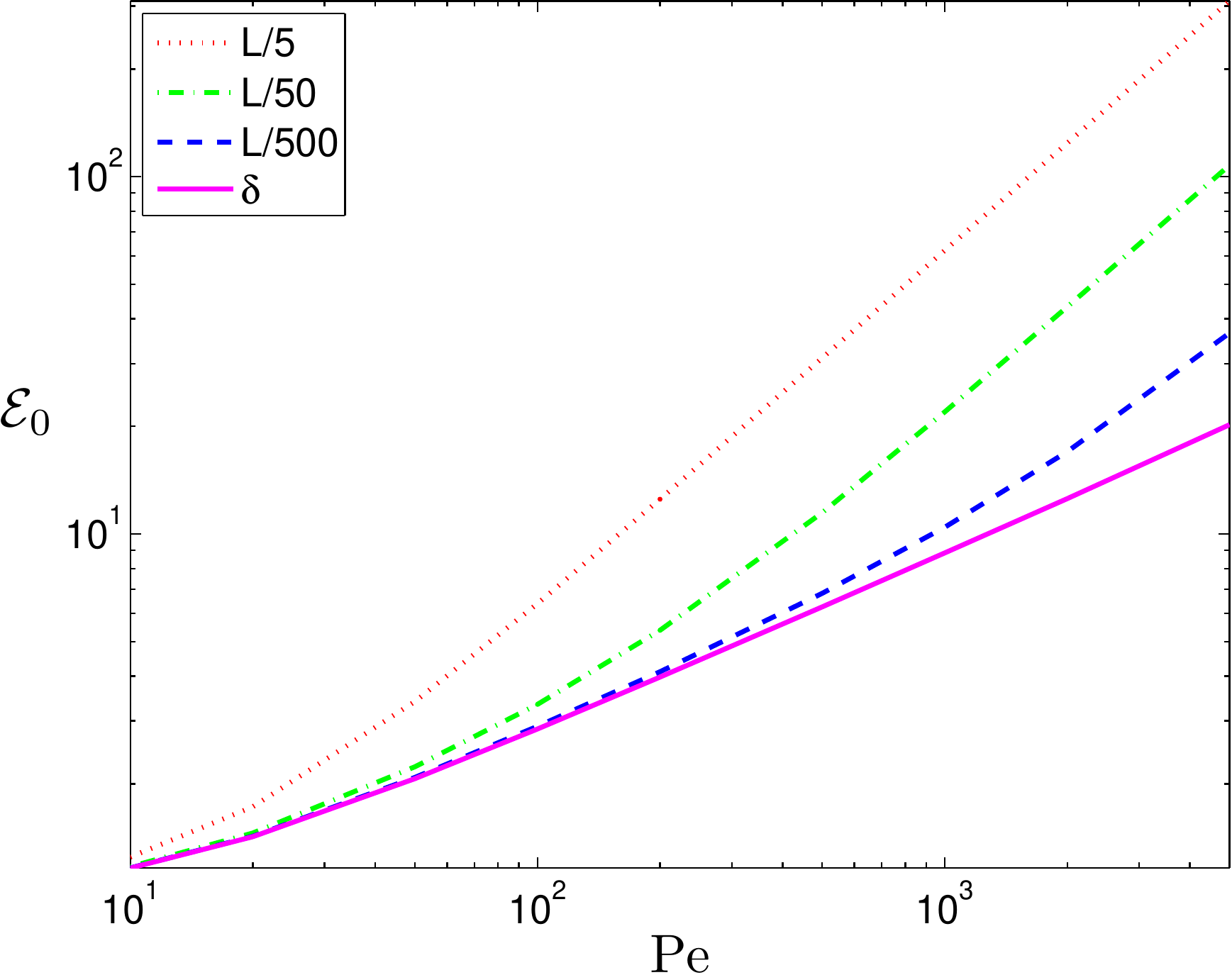}
  \label{fig:3Dbounds}
}
\caption{The theoretical upper bounds for (a) square sources with
  sizes $a=L/2, L/10, L/50$, and a $\delta$-function source; (b) cubic
  sources with sizes $a=L/5, L/50, L/500$, and a $\delta$-function
  source (from top to bottom).}
\end{center}
\end{figure}

Fig~\ref{fig:2Dbounds} shows the upper bounds on ${\cal E}_{0}$ for
square sources and a $\delta$-function source in 2D computed from (\ref{bound}).
The upper bound for any finite-size source is asymptotically $\sim $ Pe, but for the
$\delta$-function source it is $\sim {\rm Pe}/\ln {\rm Pe}$ in the large Pe limit.  
In 3D, the distinction between cubic sources and a $\delta$-function source is more
apparent as shown in Fig~\ref{fig:3Dbounds}: the upper bound for a
$\delta$-function source behaves $\sim \sqrt{{\rm Pe}}$ in 3D.
We stress that these mixing enhancement bounds apply for {\it any} statistically
homogeneous and isotropic flows stirring sources with these shapes.

\begin{figure}
\begin{center}
\subfigure[\ {$d=2$}]{
  \includegraphics[height=6cm]{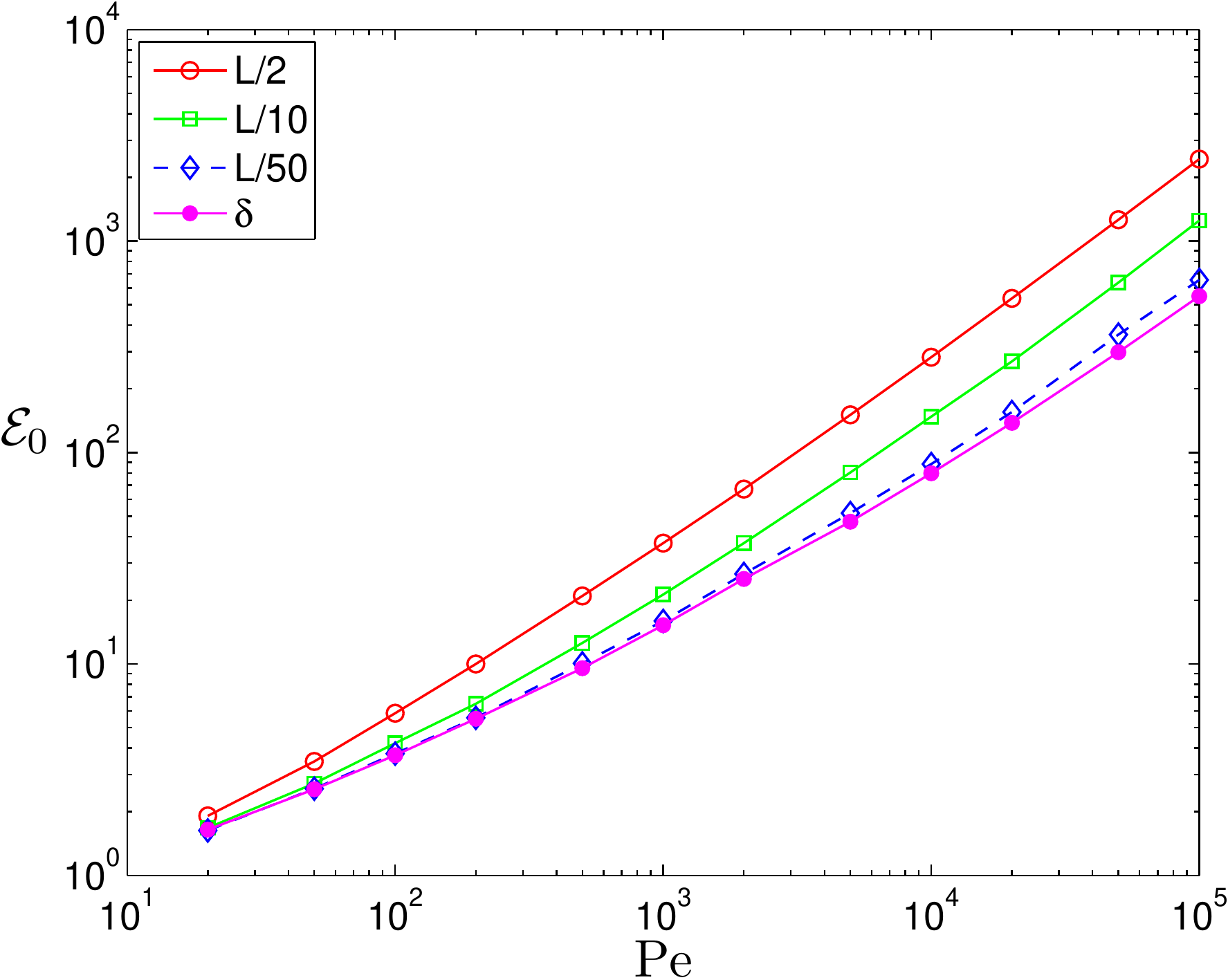}
  \label{fig:2DMixing}
}\hspace{2em}%
\subfigure[\ {$d=3$}]{
  \includegraphics[height=6cm]{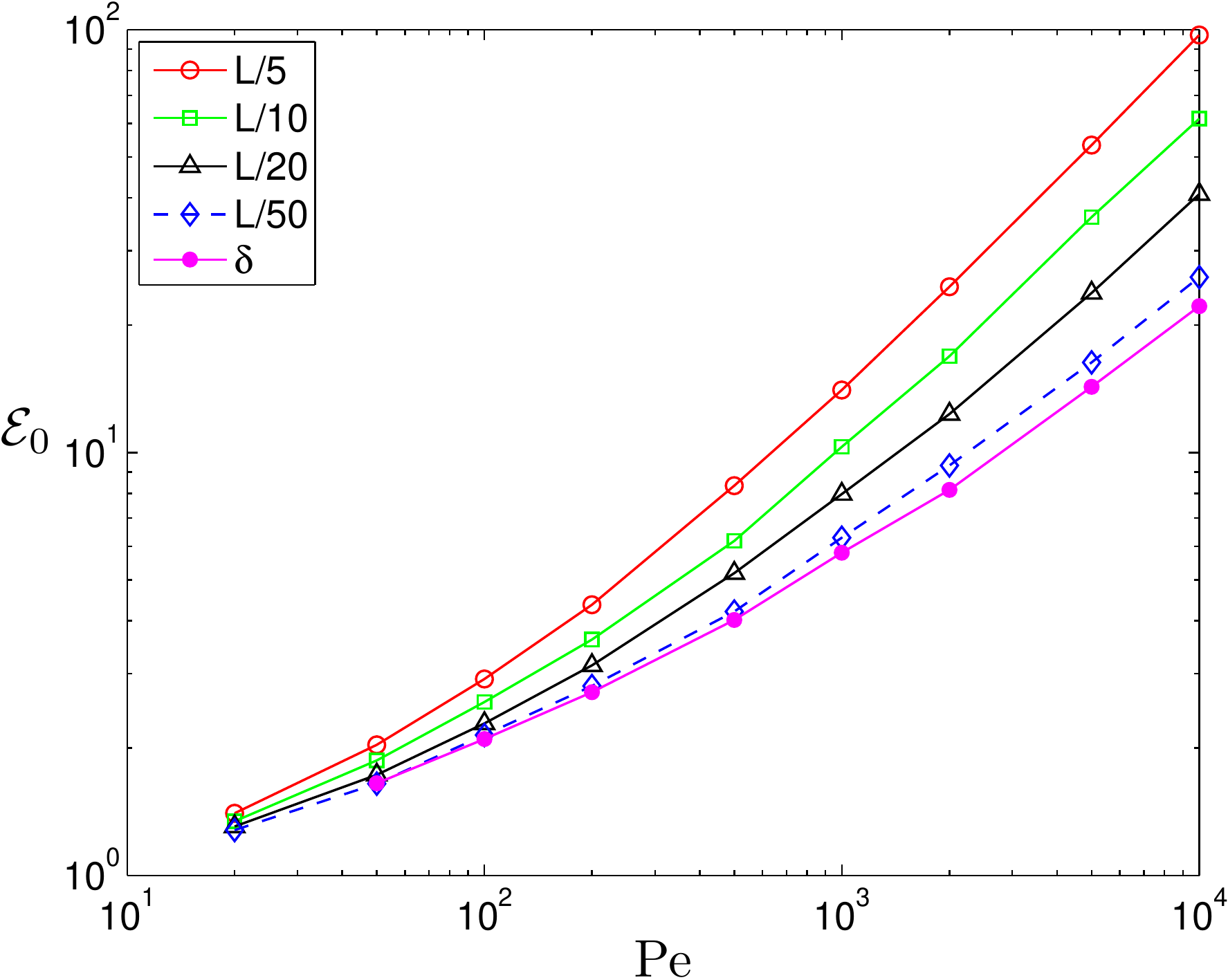}
  \label{fig:3DMixing}
}
\caption{Measured mixing enhancements for (a) square sources with sizes
  $a=L/2, L/10, L/50$ and a $\delta$-function source, (b) cubic
  sources with sizes $a=L/5, L/50, L/500$ and a $\delta$-function
  source (from top to bottom). }
\end{center}
\end{figure}

Simulation results for the random sine flow, shown in Fig.
\ref{fig:2DMixing} for 2D and Fig. \ref{fig:3DMixing} for 3D
qualitatively confirm the behavior of the enhancements suggested by the
upper limits.  
As the source size shrinks, the measured mixing enhancement gets smaller
in a way that is remarkably similar to the bounds.
In these simulations Pe is varied by decreasing $\kappa$ at a
fixed values of $L, U$ and $T$.  Other values of $T$ and other
(shorter) wavelengths of the stirring flow were also checked,
producing similar plots.  These 2D simulation results have recently been
confirmed quantitatively by a PDE computation~\cite{Chertock2008}.

\begin{figure}
\begin{center}
\includegraphics[height=6cm]{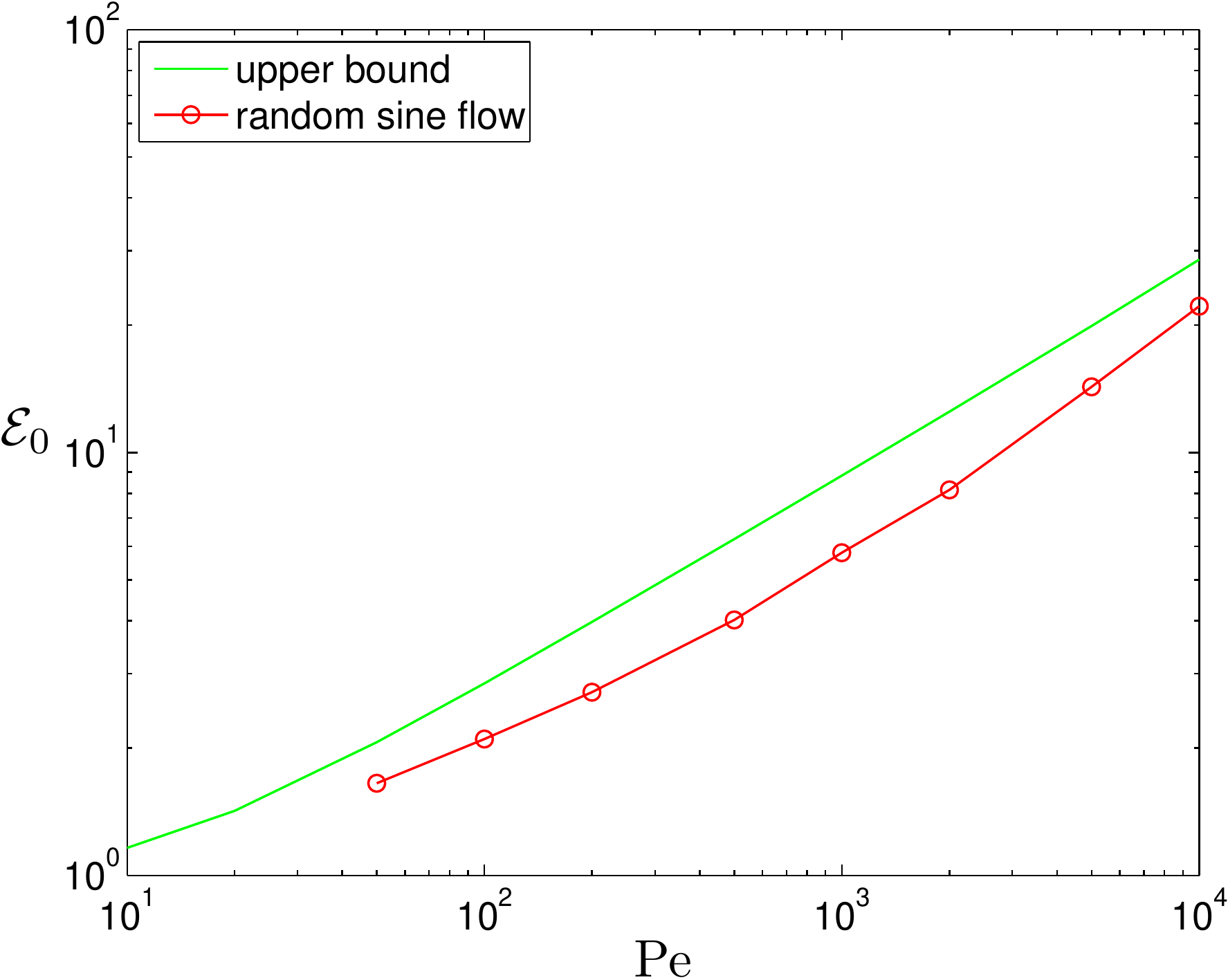}
\caption{Mixing enhancement for a $\delta$-function source.  The solid
  line is the upper bound for any SHI flow and the data are mixing
  enhancements for the random sine flow measured in the discrete
  particle simulations.}
\label{fig:3DDelta}
\end{center}
\end{figure}

The simulations also show that the upper estimates can give the
correct {\it quantitative} behavior of ${\cal E}_{0}$ as a function of
Pe.  Indeed, in Fig \ref{fig:3DDelta} we plot the upper bound on
${\cal E}_{0}$ for the $\delta$-function source in 3D and the measured
enhancement from the simulations.  The upper bound, which scales
anomalously $\sim \sqrt{{\rm Pe}}$, is an excellent predictor of the
data.  From this we conclude that the random sine flow is an
``almost-optimal'' mixer (among statistically homogeneous and isotropic
flows) for this source-sink distribution.

\section{Summary and conclusions}

We have devised an accurate and computationally efficient particle
method to study hydrodynamic variance suppression by a mixing flow.
Rigorous upper bounds for the mixing enhancement ${\cal E}_{0}$, i.e.,
the effective diffusion enhancement factor, were compared to measured
enhancements for the simple random sine flow.  A key prediction of the
analysis in Refs.~\cite{DoeringThiffeault2006,Shaw2007} is that the source-sink shape is a
determining factor in the mixing enhancement of any flow.  The simulation
results reported here show that the upper estimates give the correct
qualitative picture as regards the Pe and source-shape dependence of
${\cal E}_{0}$.

Future work should focus on investigating enhancements of other stirring
flows.  No attempt has been made here to find a more efficient
stirring flow (or, indeed, the {\it most} efficient flow, if there
is one) whose enhancement approaches more closely (or perhaps even
saturates) the upper bound.  It is remarkable that the simple
random sine flow appears to saturate the upper bound scaling
${\cal  E}_{0} \sim \sqrt{{\rm Pe}}$ in 3D.

Stirring with appropriate turbulent solutions to the
Navier--Stokes equation is also of significant interest.
The central question here is, is statistically homogeneous and isotropic
turbulence generically an efficient mixer?
The answer may depend on the source-sink distribution.

\section*{Acknowledgements}

The authors thank Jai Sukhatme for helpful comments on the paper.
This work was supported in part by US National Science Foundation
through awards PHY-0555324 and DMS-0553487, by the Geophysical Fluid
Dynamics Program at Woods Hole Oceanographic Institution, and by the
Alexander von Humboldt Foundation.


\end{document}